\documentclass[authoryear,letterpaper]{elsarticle}
\usepackage[utf8]{inputenc}
\usepackage{graphicx}
\usepackage{setspace}
\usepackage{lineno}
\usepackage[letterpaper]{geometry}
\usepackage{multicol}
\usepackage[authoryear,round]{natbib}
\usepackage{booktabs}
\usepackage{titlesec}
\usepackage{pdflscape}
\usepackage{color}
\usepackage{amssymb}
\usepackage{longtable}
\usepackage{array, multirow, bigdelim, makecell} 
\usepackage{multicol}
\usepackage{multirow}
\usepackage[normalem]{ulem}

\begin{document}

%Define the geometry
\newgeometry{left=4.5cm, right=4.5cm, bottom=4.5cm, top=4.5cm}
\graphicspath{ {figures/} }

\title{Modeling transmission windows in Titan's lower troposphere: Implications for infrared spectrometers aboard future aerial and surface missions}

\author[1,2]{Paul Corlies\corref{cor1}\fnref{fn1}}
\ead{pcorlies@astro.cornell.edu}
\author[3,4]{George D. McDonald \fnref{fn1}}
\author[1]{Alexander G. Hayes}
\author[3]{James J. Wray}
\author[5]{M\'{a}t\'{e} \'{A}d\'{a}mkovics\fnref{fn2}}
\author[6]{Michael J. Malaska}
\author[6]{Morgan L. Cable}
\author[6]{Jason D. Hofgartner}
\author[7]{Sarah M. H\"{o}rst}
\author[8]{Lucas R. Liuzzo}
\author[3]{Jacob J. Buffo}
\author[9]{Ralph D. Lorenz}
\author[9]{Elizabeth P. Turtle}

\cortext[cor1]{Corresponding author}
\fntext[fn1]{Authors contributed equally.}

\address[1]{Department of Astronomy, Cornell University, Ithaca, NY 14853}

\address[2]{Department of Earth, Atmospheric, and Planetary Sciences, Massachusetts Institute of Technology, Cambridge, MA 02139}

\address[3]{School of Earth and Atmospheric Sciences, Georgia Institute of Technology, Atlanta, GA 30308}

\address[4]{Department of Earth and Planetary Sciences, Rutgers, The State University of New Jersey, Piscataway, NJ}

\address[5]{Department of Physics and Astronomy, Clemson University, Clemson, SC 29634}
\fntext[fn2]{Lockheed Martin Advanced Technology Center\\
Palo Alto, CA 94304, USA}

\address[6]{Jet Propulsion Laboratory, California Institute of Technology, Pasadena CA, 91109}

\address[7]{Morton K. Blaustein Department of Earth and Planetary Sciences, Johns Hopkins University, Baltimore, MD 21218}

\address[8]{Space Sciences Laboratory, University of California, Berkeley, CA 94720}

\address[9]{Johns Hopkins University Applied Physics Laboratory, Laurel, MD 20723}

\begin{abstract}

From orbit, the visibility of Titan's surface is limited to a handful of narrow spectral windows in the near-infrared (near-IR), primarily from the absorption of methane gas.  This has limited the ability to identify specific compounds on the surface---to date Titan's bulk surface composition remains unknown. Further, understanding of the surface composition would provide insight into geologic processes, photochemical production and evolution, and the biological potential of Titan's surface.  One approach to obtain wider spectral coverage with which to study Titan's surface is by decreasing the integrated column of absorbers (primarily methane) and scatterers between the observer and the surface.  This is only possible if future missions operate at lower altitudes in Titan's atmosphere.  Herein, we use a radiative transfer model to measure in detail the absorption through Titan's atmosphere from different mission altitudes, and consider the impacts this would have for interpreting reflectance measurements of Titan's surface. Over our modeled spectral range of 0.4 -- 10 $\mu$m, we find that increases in the width of the transmission windows as large as 317\% can be obtained for missions performing remote observations at the surface.  However, any appreciable widening of the windows requires onboard illumination.  Further, we make note of possible surface compounds that are not currently observable from orbit, but could be identified using the wider windows at low altitudes. These range from simple nitriles such as cyanoacetylene, to building blocks of amino acids such as urea.  Finally, we discuss the implications that the identifications of these compounds would have for Titan science.

\end{abstract}

\maketitle

\section{Introduction}

Titan's atmosphere consists of a combination of methane (1.5\% -- 5.6\% mole fraction) and other higher order hydrocarbons that absorb much of the incident insolation in the near-infrared (near-IR) \citep{Kuiper1944,Flasar2005,Niemann2010}. The result of this for spectral studies of Titan’s surface in the infrared is spectral windowing--from the top of the atmosphere; Titan's surface can only be studied at a handful of methane transmission windows (referred to hereon simply as windows, \citealt{McCord2008}).

The Visual and Infrared Mapping Spectrometer (VIMS) aboard the Cassini mission put constraints on Titan's surface composition through the measurement of reflected sunlight within these windows. There are eight atmospheric windows in VIMS's 0.88 -- 5.12 $\mu$m infrared range, centered at 0.94, 1.08, 1.28, 1.6, 2.0, 2.7, 2.8 and 5.0 $\mu$m \citep{Sotin2005, Baines2005, Soderblom2010}, which can be seen as areas of higher transmission (e.g. white areas in \textbf{Fig. 1a}). Detecting specific surface compounds through reflectance spectroscopy has been difficult because of the narrow spectral coverage through the windows in Titan's near-IR spectrum. The exceptions are the reported detections of ethane in the south polar Ontario Lacus \citep{Brown2008}, of benzene and possibly cyanoacetylene \citep{Clark2010}, and acetylene \citep{Singh2016}.

Because of this spectral windowing, classification of Titan's surface composition to date has largely consisted of ``multispectral'' analysis consisting of measuring spectral slopes and band ratios between the windows.  This analysis has lead to the identification of a few spectrally distinct terrain types on Titan, including the ``5.0 micron bright,'' ``dark blue'' and ``dark brown'' regions, which are inferred to vary in their relative content of water ice vs organic compounds. \citep{Barnes2005,Rodriguez2006,LeMouelic2008,Barnes2008,Soderblom2007a,Rodriguez2014,Solomonidou2014,Rannou2015,MacKenzie2016,Brossier2018}. Additional measurements of the surface composition have included near-surface visible and near-infrared spectra taken by the Descent Imager/Spectral Radiometer (DISR) aboard Huygens, which have been interpreted as evidence for water ice and tholins \citep{Tomasko2005}. In-situ measurements by the Huygens Gas Chromatograph Mass Spectrometer (GCMS) indicated methane moisture in the ground, the detection of ethane, and possibly cyanogen, acetylene, carbon dioxide, and benzene \citep{Niemann2010}.  Using RADAR, compositional inferences based on dielectric constants have suggested the presence of H$_{2}$O in some locations \citep{Janssen2016}.  Finally, measurements of the loss tangents of the polar lakes and seas using the RADAR altimeter suggest ternary compositions of primarily methane, ethane, and nitrogen \citep{Mastrogiuseppe2014,Mitchell2015,Mastrogiuseppe2018}. Despite this large set of measurements, detections of surface compounds are restricted to a small number of possible aliphatic and aromatic hydrocarbons: methane, ethane, benzene, cyanogen, and cyanoacetylene. The primary constituents of the majority of Titan's surface remain unknown, limiting our understanding of the geologic processes operating on Titan's surface, the end products of its photochemistry, as well its potential for prebiotic chemistry.

Because of Titan's narrow near-IR windows, the majority of open questions concerning Titan's surface composition cannot be answered using existing data alone and will require future missions. In this manuscript, we quantify the transmission through and shape of Titan's near-IR windows as a function of altitude insofar as they influence observations of Titan's surface from the visible to mid-IR.  In addition, we discuss the optimal altitudes and capabilities required for future mission concepts in order to maximize the scientific return of these missions. We also demonstrate the degree to which additional light sources, such as an infrared lamp, can facilitate the use of these widened windows. Finally, we evaluate the utility of these windows for hyperspectral detections of compounds relevant for characterizing Titan's surface geologic and chemical processes.

\section{Methods}
\label{method}

The following sections detail the various inputs and parameterizations that are used in the PyDISORT radiative transfer code to model Titan's atmosphere over the spectral range of 0.4 -- 10 $\mu$m.  These inputs have been validated against VIMS observations, which cover wavelengths from 0.88-5.12 $\mu$m and the reader is directed to the Supplementary Material for more details.  For visible wavelengths, we adopt the default single scattering albedo and haze profiles from \cite{Doose2016}, but make no attempt at further validation.

\subsection{Radiative Transfer Model}
\label{method:model}

The radiative properties of Titan's atmosphere are modelled using PyDISORT, a discrete-ordinates, plane-parallel radiative transfer model, with a pseudo-spherical correction \citep{Adamkovics2016}. We utilize measurements from the Huygens probe to specify the atmospheric structure, composition, and scattering properties of aerosols (for multiple scattering) in Titan's atmosphere \citep{Fulchignoni2005,Tomasko2008,Niemann2010}.  The gas opacities used in the model come from two primary sources.  First, for wavenumbers 1000 -- 7150 cm$^{-1}$ (1.4 -- 10 $\mu$m), correlated k-coefficients are generated from line-by-line calculations of line assignments from the newly updated HITRAN2016 database \citep{Gordon2017}, with a constant spectral resolution of 4 cm$^{-1}$.  Second, for wavenumbers 7150 -- 25000 cm$^{-1}$ (0.4 -- 1.4 $\mu$m), the empirically derived k-coefficients from \cite{Karkoschka2010} are used.  These k-coefficients are generated with a constant resolution of 10 cm$^{-1}$ from 7150 -- 19300 cm$^{-1}$ (0.52 -- 1.4 $ \mu$m) and 50 cm$^{-1}$ from 19300 -- 25000 cm$^{-1}$ (0.4 -- 0.52 $\mu$m).  Combined, these linelists are found to have good agreement with the recent TheoReTS line lists over the spectral range of interest (see \cite{Rey2017} Fig. 13 and also the Supplementary Material, Fig. S3.). 

\subsubsection{Modelled Atmosphere Structure}

To assess the variation in transmission with altitude, the modelled atmosphere is divided into a minimum of 60 layers. These layers are linearly spaced in altitude, covering the range from the surface to the nominal mission altitude of interest.  Linear spacing in altitude was used for uniform altitude sampling, but has no significant impact on derived values.  Values for the temperature, pressure, and densities as measured by the Huygens probe \citep{Fulchignoni2005,Niemann2010} are then interpolated to the model layers. For simplicity, all viewing geometries are assumed to be at a zero degree incidence and emission angle (zero phase). Therefore, the results that follow are a best case scenario, as significant increase in the phase angle would increase the path length through Titan's atmosphere, increasing the effects of both absorption and scattering through Titan's atmosphere.

\subsubsection{Aerosols}

The DISR instrument on Huygens measured the aerosol extinction from 0.4$ \mu$m -- 1.6$ \mu$m and placed constraints on the monomer size and number of monomers from which the aerosols are composed as well as their fractal dimension.  From these measurements, aerosol scattering phase functions were calculated for the hazes over the range from 0.8 $\mu$m -- 5.2 $\mu$m for above and below 80 km \citep{Tomasko2008}.  Here, we  use the most recent model of Titan's aerosols developed by \cite{Doose2016} that provide improvements to the phase function and extinction of Titan's aerosols.  Therein, they adapt a single, modified phase function that is a weighted combination of the upper atmosphere (i.e. $>$80 km) and lower atmosphere (i.e. $<$80 km) phase functions derived in \cite{Tomasko2008}. For the scattering phase function, the values measured for the longest wavelength (5.166 $\mu$m) are used from 5.166 -- 10 $\mu$m.  Because the aerosol properties are so poorly constrained (even at shorter wavelengths) a more complicated extrapolation cannot be motivated.

The aerosol extinction measured by DISR was reported in \cite{Tomasko2008} and updated in \cite{Doose2016}.  Extinction curves were measured as a function of altitude for two regions: $>$55 km and $<$55 km, which have been adopted for this work. For the extinction, the power laws defined in \citep{Doose2016} are extrapolated to wavelengths out to 10 $\mu$m.   

The single scattering albedo for the hazes were only derived in \cite{Tomasko2008} and \cite{Doose2016} for wavelength up to 1.6 $\mu$m---limited by the spectral coverage of the DISR instrument.  Many previous efforts have sought to measure the wavelength dependence of the single scattering albedo of Titan's aerosols.  Given the use of the new aerosols scattering and extinction properties of \cite{Doose2016}, we have re-derived the wavelength dependence single scattering albedo for Titan's hazes as well as validated our radiative transfer model through a comparison to a spectrum extracted from the Huygens landing site with the Visual and Infrared Mapping Spectrometer (VIMS) instrument aboard the Cassini spacecraft.  For more details on the derivation of the wavelength-dependent single scattering albedo, the reader is referred to the Supplementary Material.

\subsubsection{Composition and Opacity}
\label{method:opacity}

Included in the model are N$_{2}$, H$_{2}$, CH$_{4}$, CH$_{3}$D, C$_{2}$H$_{2}$, C$_{2}$H$_{6}$, and CO abundances as measured by the GCMS instrument on the Huygens probe from an altitude of 147 km down to Titan's surface \citep{Niemann2010}.  Methane was measured directly by the GCMS instrument and was found to follow a saturation curve in Titan's troposphere down to $\sim$7 km, where the mole fraction remains constant to the surface at $\sim$5.7\%.  The CH$_{3}$D/CH$_{4}$ ratio is set to the value of 6 $\times$ 10$^{-4}$ \citep{Thelen2019}. H$_{2}$, C$_{2}$H$_{6}$, and C$_{2}$H$_{2}$ were also measured with the GCMS and found to be constant with altitude with mole fractions of 1 $\times$ 10$^{-3}$, 1 $\times$ 10$^{-5}$, and 1 $\times$ 10$^{-6}$ respectively \citep{Niemann2010}.  We use a CO abundance of 4.5 $\times$ 10$^{-5}$ based on Cassini/CIRS measurements \citep{DeKok2007}. 

Opacities are calculated using high resolution (0.001 cm$^{-1}$) line-by-line calculations to properly resolve the line shape of each transition.  Absorption spectra are generated for a grid of temperatures and pressures relevant to Titan conditions---the grid is then used to generate correlated-\textit{k} coefficients at a nominal resolution of 4 cm$^{-1}$ \citep{Lacis1991}.  Pressure and Doppler broadening effects are modelled through a Voigt profile, in combination with a far wing correction factor as described in \citep{DeBergh2012}.  We note that these correction factors have been modelled only for the 1.6 $\mu$m window, but have been found to generally agree with the other methane windows $<$ 5.0 $\mu$m \citep{Hirtzig2013}, with only a small correction for the 2.0 $\mu$m window.  Here, we do not apply the additional correction at the 2.0 $\mu$m window, as high precision model fitting to observational data is not the goal of this work.  Due to a lack of observations to further constrain the far wing line behaviour at longer wavelengths, we apply the same cutoffs to wavelengths $>$ 5.0 $\mu$m.  

Collision-induced absorptions (CIAs), resulting from the deformation of colliding non-polar molecules, are important in Titan's atmosphere for N$_2$--N$_2$ and N$_2$--H$_2$ collisions. These affect in particular the structure of the 2.0 $\mu$m window, the 2.7 $\mu$m and 2.9 $\mu$m doublet, and the short-wavelength wing of the 5.0 $\mu$m window \citep{Soderblom2010}. Absorption coefficients for the CIAs measured at Titan relevant conditions are taken from \citealt{McKellar1989} and \citealt{Hartmann2017}.  While we make use of all the most recent models for CIAs, we note that the center of the 2.0 $\mu$m window is still measured at longer wavelengths than what is observed with VIMS (the equivalent of 1-2 VIMS channels), which could suggest vertical variations in abundance or temperature that are not fully captured with the model.  

The final component of the model is the parameterization of the surface scattering, which is only used in the calculation of apparent reflectance in \textbf{Fig. 2}. In this work, we assume a lambertian surface scattering with a grey albedo of 0.1 for all wavelengths. Although we know the spectral variation of Titan's surface within the narrow windows available from orbit, the albedo outside of these regions (over much of the near-IR) is still unknown (and due to specific surface compound absorption features it likely cannot be interpolated between the windows---i.e. it is very possibly not monotonically increasing or decreasing as a function of wavelength).  Therefore, we opt for the simple model of a grey scattering surface for this work---this best keeps with our focus on quantifying the effects that the atmosphere in particular has on observations of the surface. 

\subsection{On the effect of hazes}
\label{method:hazes}

Hazes play an important role in our calculations as they both absorb and scatter incident radiation. The focus throughout the paper is on the effects of gaseous and haze absorption as these are what primarily affect the shapes and utility of Titan's near-IR windows. We are thus reporting what wavelengths \textit{could} be of utility as absorption from atmospheric gases has been significantly reduced or eliminated. However, we include an approximation for the loss of transmission \cite{Haberle1993} from Rayleigh and haze scattering to demonstrate to the reader the relative contributions of these terms.  Further, scattering is fully included in the calculations of apparent reflectance (i.e. the results in \textbf{Fig. 2}).  Scattering plays a complicated role in determining the diffuse illumination and is thus important for considerations of the spatial resolution of a given instrument/mission (e.g. \cite{Barnes2018,Barnes2020}).  The \textit{practicality} of each spectral window will depend on the relative importance of scattering, which depends on the specific mission architecture and scientific objectives.

\subsection{Surface compounds of interest}
\label{method:surf_comp}

Our chief interest is in using the widened atmospheric windows available in the lower atmosphere for studies of Titan's surface composition. Although examining a comprehensive list of all possible compounds of interest would not be possible within the scope of this manuscript, we focus on several categories of compounds: 

\begin{enumerate}
    \item Hydrocarbons and nitriles. These include primary products predicted from photochemical models as well as those molecules that would inform the primary photochemical pathways \citep{Lavvas2008a,Lavvas2008b,Krasnopolsky2009,Vuitton2019}. They also include those compounds that have been directly measured in Titan's atmosphere \citep{Vinatier2015,Horst2018} and could be accumulating on the surface.
    \item Simple inorganic compounds that could comprise the bedrock of Titan's surface or indicate communication from the interior to the surface.
    \item Nucleobases, amides, and amino acids, which would be indicative of the prebiotic potential of Titan's surface. Here our list includes those that have been measured in either meteorites \citep{Burton2015,Cronin1983,Wolman1972} or cometary material \citep{Hadraoui2019,Elsila2009}, and those which are of particular relevance to Titan as discussed in section \ref{disc}.
\end{enumerate}

Over the 0.4 -- 10 $\mu$m spectral range that we model for the atmosphere, we selected the center of spectral bands for the above compounds. We identified which features are measured to be ``strong'', taken to consist of a transmission $<$30\% for the given method.  When possible, we note which features are considered to be ``wide''. Many bands, particularly the strong ones focused on in this manuscript, merge, presenting themselves as one large band depending on: (1) the spectral resolution of the spectrometer used for the observations and (2) the amount of the material present. Therefore, the ''width'' is simply measured as the width of the absorption feature at 90\% of the peak transmission, and is considered to be wide if this is $>$10 cm$^{-1}$. This criteria generally avoids the problem of merged lines, while simultaneously avoiding the misidentification of narrowband absorptions. The reader is cautioned to take the determination of ``wide'' features with these caveats into consideration. The reader is also referenced to each of the source databases, described below, which can be used for ad-hoc determination of a particular band strength and width. Further, \cite{Clark2010} provides some discussion for the width of several absorptions features in the 5.0 $\mu$m window, to which the reader is also referenced.

For this study, four primary spectral databases were used for compiling the various compounds' absorptions.  First, is the Spectral Database for Organic Compounds \citep{SDBS2020}, which provides a majority of the absorptions for nucleobases and amino acids.  This database consists primarily of transmission spectra through 150 $\mu$m samples, though the thickness of the sample is varied to optimize transmission.  The second database is the Cosmic Ice Laboratory optical constants, which provides the optical constants for various species under various temperatures and types of ice structure (amorphous, crystalline, metastable) \citep{Hudson2020}.  Here, we use the crystalline optical constants at the temperature closest to 95 K, to best match Titan surface conditions.  We note that methane ice is only available at 10 K.  From these constants, simulated transmission spectra are generated assuming a sample thickness of 1.0 $\mu$m.  The third database is provided from the NIST Chemistry WebBook, which offers a variety of spectra (gases, liquid, solution) for various compounds in the near-IR  \citep{NIST2020}.  Here, we choose compounds that are available in liquid or solution.  The only exception is for biphenyl (C$_{12}$H$_{10}$), for which only a gaseous spectrum is available.  However, we find good agreement between this spectrum and the one provided in the SDBS database. The source is the Ice Analogs Database, which offers absorbance spectra for various inorganic ices at 10 K, with sample thickness varying between 0.11 -- 0.15 $\mu$m \cite{Gerakines1995}.  We convert absorbance to transmission and use these values for determining absorption bands.  These four databases typically run over the spectral region from 1000 -- 4000 cm$^{-1}$ (2.5 -- 10 $\mu$m).  Finally, we have supplemented our chosen compounds with values found in the wider literature.  For C$_{2}$H$_{2}$, additional absorptions for wavenumbers $>$4000 cm$^{-1}$ ($<$2.5$\mu$m) are taken from \cite{Singh2016}.  For all alkanes, additional absorptions for wavenumbers $>$4000 cm$^{-1}$ ($<$2.5$\mu$m) are taken from \cite{Clark2009}.  \cite{Clark2010} also provides a table of absorption features for various Titan-relevant compounds in the 5.0 $\mu$m window. When appropriate, those values have also been included here.

\section{Results}

\subsection{Transmission vs Altitude}
\label{res:trans_vs_alt}

When calculating transmission, as in \textbf{Fig. 1}, we consider only the radiation that is directly absorbed by the hazes, which can be modelled through $\tau_{A} = \tau (1-\omega_{o})$, where $\tau$ is the total extinction as measured by DISR.  Combining this haze absorption with gaseous absorption, we calculate the two-way transmission through Titan's atmosphere defined as \begin{equation} T=e^{-2\tau_{\lambda,z}} \end{equation} where $\tau$, the integrated one way opacity, is a function of wavelength and the altitude of the observation.  We focus here on the windows $>$0.88 $\mu$m, which are most useful for uniquely identifying surface compounds, and which have been validated against VIMS observations (see Supplementary Material). For completeness, we include windows down to visible wavelengths including a full radiative transfer treatment. However, validation of these wavelengths (i.e. comparison to Huygens data) was not performed, as they are not as practical as their near-IR counterpart for uniquely identifying surface compounds, and are therefore not the primary focus of this work.

\textbf{Fig. 1} plots the expected transmission resulting from the absorption of the primary components of Titan's atmosphere for four characteristic mission altitudes: lander, drone, balloon, and orbiter.  Methane acts as the primary absorber in Titan's atmosphere and generates most of the features observed in Titan's near-IR transmission.  We find that most wavelengths, besides those that are transparent from orbit, remain relatively opaque unless close to the surface, demonstrating that only a reduction in the integrated column of methane can appreciably improve the near-IR visibility of Titan's surface.  Only at altitudes less than $\sim$1 km is there any appreciable change in the effects from methane absorption.  For example, the 5.0 $\mu$m window width is increased by only $\sim$5\% relative to orbit at an altitude of 1 km, but this broadening increases to 17\%, and $\sim$220\% for altitudes of 100 m and 1 m, respectively (see \textbf{Table 1} for reported window widths based on percent transmission).  This is the most dramatic broadening of Titan's windows, but demonstrates the extent to which future missions need to operate as low as possible in Titan's atmosphere in order to increase the portions of the spectral domain that are available for studies of surface compositions. 

For reference, we also show the effects haze and Rayleigh scattering (using the approximation of \cite{Haberle1993}) were they to be treated independently as dashed lines. While these significantly reduce the transmission through the atmosphere from orbit, their smoothly varying spectral behavior has little effect on the \textit{shape} of the windows. Furthermore, from 1 km and below, the contributions from these scattering terms is almost negligible. For these reasons and as per the discussion in Section \ref{method:hazes}, we do not include these effects on transmission when quantifying window widths in \textbf{Table 1} and \textbf{Fig. 3}.

In addition, we find strong, but narrow absorptions from each of the trace gases, some in areas of particularly clear methane absorption.  In these cases, such as for CO and C$_{2}$H$_{2}$ at 5.0 $\mu$m, these narrow absorptions help to define the width of various windows in Titan's near-IR spectrum.  We note that C$_{2}$H$_{6}$ (ethane) absorptions are located primarily in strong methane absorption bands, though some recent work suggests that C$_{2}$H$_{6}$ may also contribute to the shapes of the 2.0 $\mu$m and 2.7 $\mu$m windows \citep{Maltagliati2015,Rannou2018}.  Considering this, and given the role that ethane plays in Titan's hydrology (e.g. clouds, composition of lakes, etc.) and atmospheric chemistry, we have included it for completeness.

For the orbital case, we find broad absorption resulting from Titan's hazes, especially at shorter wavelengths where the aerosol opacity is higher.  Absorption from the hazes is found to drop dramatically by 1 km, with a maximal opacity at this altitude of $\tau<$.01.

We note that the HITRAN line list that we use, as well as all other currently available line lists for Titan relevant conditions, are incomplete from 2.9-3.2 $\mu$m. We caution the reader that there may be additional absorption features in this region that cannot currently be modelled (see \textbf{Fig. S2} for a comparison of our modeling of this region to existing orbital observations from VIMS).

%\sout{We note, here, that the treatment of the window widths as the direct comparison to improvement from orbit does not consider some possibly important additions.  For example, although there may not be significant improvement in the spectral coverage for low-altitude missions at $\sim$1km, there is likely to be an improvement in the detection of surface compounds resulting from the improved spatial clarity resulting from the decrease in haze column at these altitudes.  Understanding the relative percentage of light that has been directly transmitted, single-ly scattered, or multiply scattered remains a major difficulty in interpreting spectra of Titan's surface.  Of course, at longer wavelengths, where the opacity of the hazes is minimal ($>\sim$2.7 $\mu$m), this consideration becomes less imporant and the improved spectral coverage acts as the primary limiting factor in observing Titan's surface. Here, we focus on what improvement can be expected in spectral coverage alone, but these additional considerations will also be necessary when planning the development of future missions and instruments.}\textcolor{red}{Now included in section \ref{method:calc_trans}}.

\subsection{Expected Signal: Passive vs Active Illumination}
\label{res:passive_active}

Although transmission through the atmosphere can be improved by decreasing the mission altitude, it still does not mean that these spectral regions are immediately accessible. This is because solar illumination must pass through Titan's opaque atmosphere. \textbf{Fig. 2} shows spectra of Titan's reflectance at two altitudes in the atmosphere, comparing the differences between passive (solar) and active (lamp) illumination.  For these calculations of reflectance, a full scattering treatment of the hazes has been included.

Because the one-way integrated column density through the entirety of Titan's atmosphere is large, solar illumination is mostly absorbed before it reaches the surface (see \textbf{Fig. 2}a).  Only in the transparent windows can it provide appreciable signal at the surface.  However, if a mission were to include an active light source, such as a broadband lamp or LEDs, then the integrated column density from source to detector is simply the two-way distance between the spacecraft and the surface.  This is significantly shorter than in the passive observation case, in which photons must traverse the entirety of the atmosphere once plus the distance from the surface to the observation altitude (after reflection off of the surface). Thus, active observations are required to access any widened windows. We note, however, that unless the windows have widened appreciably with respect to those observed from orbit, which we demonstrate in \textbf{Fig. 2a} to \textbf{Fig. 2b}, use of a lamp will not necessarily increase spectral coverage significantly.  Thus, the use of onboard illumination is only suggested for drone or lander class missions.

There are several types of possible illumination sources that could be used.  \textbf{Fig. 2b} plots a comparison between broadband and narrowband (LED) light sources.  A broadband source offers the benefit of wide spectral coverage, but at the cost of significant power. LEDs much more efficiently emit in narrowband regions, however, it can be seen that for the wider spectral regions, particularly $>$ 2.7 $\mu$m, multiple LEDs are required to provide comparable coverage to a single broadband source (see \textbf{Fig 2b}).  LEDs are also less efficient at longer wavelengths, and thus begin to require more power in this region.  Short of 2.7 $\mu$m, although the windows do broaden significantly, they are still narrow enough such that a single, tuned LED can cover the entire spectral window.  We note that the exact spectral coverage of an LED and its ability to provide sufficient illumination will depend on both its temperature and power.  We define the available regions in \textbf{Fig. 2b} from commercial LEDs (typically with a power of a couple mW), limiting their spectral coverage to regions of 80\% peak spectral density. However, the exact signal-to-noise ratio (SNR) that can be obtained in these regions will depend on the number and power of LEDs as well as their angular distribution (which would dictate how concentrated their light is).  We do not model SNRs here given that they are highly dependent on the properties of spectrometer instrument suites (telescope size, detector, integration times etc.).  However, given an LED's ability to concentrate power within narrow spectral regions, they act as a viable, low-energy counterpart to broadband illumination for narrow regions of interest.  Thus, a probable mission architecture in the future could include a combination of LED sources for shorter wavelengths and a broadband source for longer wavelengths, each of which can be selected depending on the observations being made.

%We remind the reader that upon selection of a mission architecture, detailed modeling of the haze scattering is necessary as discussed in section \ref{method:calc_trans}. An additional consideration related to active illumination is the necessity of shielding the detector from the bright source itself, to allow for the detection of much fainter signals from a more distant surface.

%\sout{Again, we note that in this work we focus our definition of "improvement" for various mission architectures as increased spectral coverage, resulting from the decrease in the spectral column.  However, other contributions from the highly scattering hazes could play an important role, even at low altitudes, where the optical depth of the haze column is small.  For example, bright scattering from nearby haze particles could influence the ability for accurate surface detection, if that surface was sufficiently far (e.g. balloon-borne architecture). Another consideration would be shielding the detector from the bright source itself, to allow for the detection of the much fainter signal from a more distant surface.  Appropriate baffling of the instrument would therefore be required and may pose a significant challenge for a reliable detection of surface compounds, especially at shorter wavelengths where Titan's hazes have stronger scattering effects.}

\subsection{Ability to detect surface compounds}
\label{res:detection_surface}
We now discuss what potential surface compounds of interest, from the categories discussed in \ref{method:surf_comp} could be detected in the widened spectral windows in the lower atmosphere. 

\textbf{Fig. 3} shows the spectral windows available from altitudes relevant for future balloon (1 km), drone (100 m), and lander (1 m) mission architectures. Throughout this discussion make references to the ability to detect compounds from altitudes of 10 m, which we also modeled, but are not shown in \textbf{Fig. 3}. These same values are also tabulated in \textbf{Table 1}. The windows are defined as regions where the atmospheric transmission is between 90\% (minimum width) and 10\% (maximum width).  Note that in defining these, we include gaseous and haze absorption and not scattering, as per the discussion in section \ref{res:trans_vs_alt}. The absorption bands of surface compounds of potential interest are also shown, with the different subpanels showing compounds of different types. These absorption bands are also tabulated in \textbf{Table 2} as the bold values, as only the strong absorptions are shown in \textbf{Fig. 3}.  We note that similar studies for identification of organic compounds from orbit have been conducted for the 5.0 -- 6.0 $\mu$m region \cite{Lorenz2008a}, the 2.6 -- 2.8 $\mu$m and 5.0 -- 5.2 $\mu$m windows accessible to VIMS \citep{Clark2010}, but we expand on these here considerably by considering all windows between 0.4 $\mu m$ -- 10 $\mu m$, the ability to detect at several mission architecture altitudes, and a wider range of compounds.

\textbf{Fig. 3a} indicates the absorption bands of amides and amino acids. Only two compounds in our list, urea, and sarcosine, have absorptions that lie within spectral windows available from orbit (the 2.8 $\mu$m and redward edge of the 5.0 $\mu$m window, which is not covered by the VIMS detector for urea, and the 2.8 $\mu m$ window for sarcosine). Nevertheless, these absorptions lie in regions of low ($\sim$ 10\%) transmission. The significant widening of the 5.0 $\mu$m window for a drone observing from altitudes of 100 m makes the detection of all listed compounds (glycine, sarcosine, $\beta$-alanine, and $\gamma$-aminobutyric acid) a possibility. In addition, the closer one moves to the surface, the further the two absorption bands for urea (2.91 and 5.96 $\mu m$) move from the edge to within the windows. The ability to identify a compound based on two absorption bands, as for urea in this case, would significantly increase confidence in its purported detection. Observing from altitudes of less than 10 m places absorption features of additional amino acids (glycine and $\beta$-alanine) in wide enough spectral windows of high transmission that would likely enable true hyperspectral observations. 

The ability to detect nucleobases and potential alternative nucleobases is examined in \textbf{Fig. 3b}. Although multiple species have absorption features in the 5.0 $\mu m$ window from orbit (although at wavelengths redder than the 5.2 $\mu m$ limit of the VIMS detector), it is only at 1 km or better where these features fall within areas of higher transmission. For example, observing from altitudes of 100 m places multiple cytosine absorption bands in window regions of high transmission, facilitating the possibility for detection. In addition, from 100 m all other species now have one absorption feature in regions of high transmission.

For the alkanes \textbf{Fig. 3c}, many lines remain inaccessible unless observations are made from the surface at which point groups of lines at 2.4, 3.4, and 6.8 $\mu$m fall within the widened windows.

\textbf{Fig. 3d} demonstrates the prospect of detecting nitriles and other unsaturated nitrogen-containing molecules within the newly widened windows. From orbit, only single absorption bands of indole and hydrogen cyanide (HCN) fall within the transmission windows, albeit at the window edges and in areas of low transmission. Although prospects for detection of indole improve at 1 km, it is not until altitudes of $\sim$100 m or less that the 2.94 $\mu$m indole and 4.76 $\mu$m HCN absorption bands fall comfortably within areas of high transmission in the widened windows. Again from 100 m, a $\sim$3.2 $\mu$m feature shared by HCN and acrylonitrile (CH$_2$==CHCN) falls in a window region of lower transmission. From altitudes of 10 m or less, this feature falls into an area of higher transmission, in addition to the 4.48 $\mu m$ acrylonitrile absorption features.

For alkenes and unsaturated hydrocarbons, which are shown along with nitriles in \textbf{Fig. 3d}, only absorption bands for acetylene, at 4.94 $\mu$m \cite{Singh2016}, and benzene at 5.05 $\mu m$ \cite{Clark2010}, fall within a window from orbit. For altitudes of 100 m, a second acetylene absorption band at 3.10 $\mu$m can be observed and falls into area of higher transmission for altitudes of 10 m or less.  At altitudes of less than 10 m, naphthalene and biphenyl now have absorption bands in a window region with low transmission between 6.62-6.75 $\mu$m. 

The absorption bands for inorganics are shown in \textbf{Fig. 3e}. From orbit, only water ice has an absorption band within a window at 1.9 $\mu$m, however this feature is broad and thus water ice has yet to be uniquely identified on Titan's surface. Observations improve significantly by 100 m, allowing for absorption features of hydrated minerals and CO$_2$ to fall in regions of high transmission. From 1 m, all three compounds each have a minimum of 4 strong absorption features within areas of high transmission.

We note that although increased spectral coverage allows for improved identification of these species, it is unlikely that the surface is composed of a single constituent and is likely a mix of some or many of these species.  This mixing will affect the ability to make unique characterizations of the surface composition.  However, the ability to search for and potentially identify these species still requires the increased spectral coverage as described above.

\section{Discussion}
\label{disc}

We discuss here some of the major questions in Titan surface and atmospheric science that could be addressed through the identification of compounds in these wider windows.  The bulk composition of Titan's surface remains unknown. Identifying water ice vs benzene, acetylene, or simple nitrile compounds such as indole as major surface constituents would have implications for the geologic processes operating on Titan's surface and the chemical processes occurring in Titan's atmosphere. For example, karstic evolution has been proposed as a formation mechanism for Titan's polar lakes, but requires a surface material that is soluble in liquid hydrocarbon \citep{Hayes2008,Lorenz1996,Malaska2011,Malaska2014,Cornet2015,Michaelides2016}. Testing this formation model ultimately requires that soluble surface materials are identified. Similarly, the evaporite hypothesis for the abundant 5.0 $\mu$m bright material on Titan's surface also requires soluble material(s) and identifying their composition would significantly improve the understanding of Titan's hydrology. Further, the ability to discriminate between hydrocarbons and water ice using the latter's 3.29 $\mu$m feature would help to determine what the bulk composition of the dunes is \citep{Barnes2008,Hayne2014,Rodriguez2014,Brossier2018,Solomonidou2018}, and further constrain differences between the dunes and interdunes related to possible grain-size sorting or material propensity for triboelectrification \citep{Bonnefoy2016,MendezHarper2017,Lucas2019}. With regards to aeolian processes, identifying areas of high methane or ethane content in concert with degraded dune fields could indicate that moisture-related cohesion is responsible for the inactivity of these dunes \citep{Ewing2015,McDonald2016}. Finally, candidate cryovolcanic features have been identified, including the $\sim$1 km high mountain adjacent to a $\sim$1.5 km deep depression at Sotra Facula \citep{Lopes2013}. The observation of hydrated silicates around these features could be indicative of communication with Titan's deeper interior and strengthen the case for their cryovolcanic origin. The ability to discriminate between hydrated silicates and pure water ice is only possible with broader spectral windows at low altitudes.

The identification of specific compounds on the surface can also inform our understanding of Titan's photochemistry. An area of debate has been whether aromatic rings are fused (forming polyaromatic hydrocarbons, with naphthalene being the smallest of these molecules, followed by larger molecules including phenanthrene and anthracene) or linked (forming polyphenyls) during chemical synthesis in Titan's atmosphere \citep{Delitsky2010,Vuitton2019}. Comparing the relative abundance of biphenyl and naphthalene (which may be possible from altitudes of 10 m) would be a major step in elucidating the dominant synthesis pathways for aromatic compounds on Titan.

Further information on the surface composition is also essential for characterizing the potential prebiotic chemistry occurring on Titan's surface. Laboratory studies of reactions between tholins and liquid water indicate that essential prebiotic building blocks including urea and amino acids can form \citep{Khare1986,Nguyen2007,Raulin2007} over timescales relevant for impact melt events \citep{Neish2008,Neish2009}. Such impactors would likely also carry a variety of amino acids themselves \citep{Burton2012}. We have demonstrated that the widened spectral windows available in the lower atmosphere could help confirm that these processes are occurring, as discussed in section \ref{res:detection_surface}.

Furthermore, \citealt{Jeilani2015} and \citealt{Horst2012} have proposed mechanisms by which urea, along with acetylene sourced from Titan's atmosphere, can form pyrimidine nucleobases. However, whether such biologically relevant compounds are actually forming on Titan's surface, what specific ones are being formed, as well as their abundances remain open questions. These questions could be addressed by the ability to identify uracil from altitudes of 1 km, as well as a host of other nucleobases from altitudes of 100 m or less (see \textbf{Fig. 3b}), including adenine which could be formed directly from atmospheric chemistry \citep{Horst2012,Jeilani2015}. The building blocks urea and acetylene being visible in the windows from altitudes of 1 km or less would also be of utility.

Finally, recent studies using molecular simulations suggest that nitrile compounds, such as acrylonitrile, measured in situ in Titan's upper atmosphere \citep{Vuitton2007} and from remote sensing\citep{Palmer2017}, may form membranes and vesicles (termed ``azotosomes'') that could provide the function that liposomes do for biological organisms on Earth \citep{Stevenson2015}, although this has been disputed by \citealt{Sandstrom2020}. An important step towards assessing whether such azotosomes exist on Titan is confirming that these nitrile compounds also exist on the surface and are not altered during transport through the atmosphere. We have shown that acrylonitrile may be detected from 100 m, and could comfortably be detected hyperspectrally from altitudes of less than 10 m.

Although the Huygens DISR instrument used active illumination in the lower atmosphere to study the surface composition, the majority of major absorption bands for organic compounds of interest discussed above lie at wavelengths greater than $\sim$2.8 $\mu$m, significantly redder than the 0.8 -- 1.6 $\mu$m wavelength range of Huygens DISR. Only a future mission with much broader spectral coverage could detect these compounds.

\section{Conclusions}

In summary:

\begin{itemize}
  \item We have demonstrated that the infrared spectral windows in the wavelength range 0.4 -- 10 $\mu$m broaden considerably within the context of infrared spectrometers operating in the lower atmosphere.  The broadening is small down to $\sim$ 1 km with only a $\sim$ 68\%  increase in total spectral coverage (an additional 1.4 $\mu m$ in width over the 9.6 $\mu m$ range considered in this study). It becomes appreciable at altitudes of 100 m or less, such that by an altitude of $\sim$ 1 m there is a $\sim$ 317\% total increase (additional 6.53 $\mu m$ width over the same range).
  \item Onboard illumination is required in order to observe spectra of the surface within the widened windows in the lower atmosphere. This is because the availability of \textit{solar} radiation in the lower atmosphere is largely restricted to within the same windows available from orbit.
  \item Enhanced spectral coverage will provide insight into Titan's history and evolution and its potential to act as a biologically viable environment. 
\end{itemize}

Several missions (JET, TIME, AVIATR, OCENAUS, etc.) have been proposed for Titan \citep{Lorenz2008a,Lorenz2008b,Lorenz2008c,Sotin2011,Barnes2012,Stofan2013,Sotin2017}, and the Dragonfly mission has been selected for a return to Titan in the 2030's \citep{Turtle2017, Lorenz2018}. Studying Titan's surface composition is an explicit goal of all of these missions---our quantification of the widened spectral windows will assist in designing instruments for future low-altitude missions.  An improved understanding of the detection of surface compounds possible by varying mission architectures for Titan, therefore, remains important for the development and planning of these missions.

\section{Acknowledgements}
PC acknowledges funding for this work by the NASA Earth and Space Science Fellowship Program: Grant Numbers NNX14AO31H S03. GDM acknowledges funding from the Department of Defense (DoD) through the National Defense Science \& Engineering Graduate Fellowship (NDSEG) Program, Fellowship 32 CFR 168a. RL acknowledges the support of NASA Grant 80NSSC18K1389. This research was also supported by the Cassini-Huygens mission, a cooperative endeavor of NASA, ESA, and ASI managed by Jet Propulsion Laboratory, California Institute of Technology under a contract with National Aeronautics and Space Administration. 

\section{Appendix A: Supplementary Material}
The supplementary material for this article includes a description of the model validation with comparisons to VIMS observations of the Huygen’s Landing Site. Further, it includes a comparison between different line lists and k-coefficients.

\newpage
\newgeometry{left=2cm, right=2cm, bottom=1.5cm, top=1.5cm}
\begin{landscape}
\section*{Tables}

\begin{tabular}{l | c || c | c || c | c || c | c }
\hline \hline
Altitude & Orbit (1500 km) & \multicolumn{2}{c||}{Balloon (1 km)} & \multicolumn{2}{c||}{Drone (100 m)} & \multicolumn{2}{c}{Lander (1 m)} \\
\hline
Transmission & 10\% & 90\% & 10\% & 90\% & 10\% & 90\% & 10\% \\
\hline \hline
          & 0.49 - 0.70 & 0.40 - 0.72\hspace{0.1em}\rdelim\}{3}{*}[ ] &             & 0.40 - 0.86\hspace{0.1em}\rdelim\}{3}{*}[ ] &             & 0.40 - 2.25\hspace{0.1em}\rdelim\}{3}{*}[ ] &            \\
          & 0.74 - 0.77 & 0.74 - 0.77\hspace{1.5em}                   & 0.40 - 0.88 & 0.90 - 0.97\hspace{1.5em}                   & 0.40 - 1.13 & 2.44 - 2.56\hspace{1.5em}                   & 0.40 - 3.22\\
          & 0.81 - 0.84 & 0.81 - 0.84\hspace{1.5em}                   &             & 1.03 - 1.11\hspace{1.5em}                   &             & 2.62 - 3.17\hspace{1.5em}                   &            \\
          & 0.92 - 0.96 & 0.92 - 0.95                                 & 0.90 - 0.98 & 1.21 - 1.31                                 & 1.19 - 1.32 &                                             &            \\
          & 1.04 - 1.10 & 1.05 - 1.09                                 & 1.02 - 1.11 & 1.49 - 1.62                                 & 1.42 - 1.63 &                                             &            \\
          & 1.25 - 1.30 & 1.25 - 1.30                                 & 1.21 - 1.31 & 1.88 - 1.90\hspace{0.1em}\rdelim\}{2}{*}[ ] & 1.82 - 2.17 &                                             &            \\
Window    & 1.53 - 1.61 & 1.53 - 1.61                                 & 1.48 - 1.62 & 1.98 - 2.14\hspace{1.5em}                   &             &                                             &            \\
ranges    & 1.98 - 2.12 &      -                                      & 1.87 - 1.91 & 2.64 - 2.99                                 & 2.62 - 3.02 &                                             &            \\
($\mu m$) & 2.66 - 2.81 & 1.99 - 2.11                                 & 1.97 - 2.15 &      -                                      & 3.06 - 3.14 &                                             &            \\
          & 2.90 - 2.95 & 2.77 - 2.81\hspace{0.1em}\rdelim\}{2}{*}[ ] & 2.63 - 3.00 &                                             &             &                                             &            \\
          &             & 2.92 - 2.95\hspace{1.5em}                   &             &                                             &             &                                             &            \\
          \cline{2-8}
          & 4.85 - 6.03 & 4.89 - 5.95                                 & 4.81 - 6.05 &      -                                      & 3.96 - 4.24 & 3.99 - 4.59\hspace{0.1em}\rdelim\}{5}{*}[ ] &            \\
          &             &                                             &             &      -                                      & 4.31 - 4.57 & 4.75 - 6.25\hspace{1.5em}                   &            \\
          &             &                                             &             & 4.82 - 6.03                                 & 4.77 - 6.15 & 6.39 - 6.48\hspace{1.5em}                   & 3.61 - 7.37\\
          &             &                                             &             &                                             &             & 6.52 - 6.64\hspace{1.5em}                   &            \\
          &             &                                             &             &                                             &             & 6.83 - 7.14\hspace{1.5em}                   &            \\
          \cline{2-8}
          & 9.96 - 10.00 &     -                                   & 9.26 - 10.0 & 9.61 - 10.0                                 & 8.82 - 10.00 & 8.16 - 10.0                                & 7.99 - 10.0 \\
\hline \hline
\end{tabular}

\textbf{Table 1}. Spectral windows defined based on 90\% or 10\% transmission through the distance from the instrument to the surface for each mission architecture as described in the parentheses. We note that 0.4 $\mu m$ and 10 $\mu m$ are the spectral boundaries of our model. The 10\% and 90\% quoted transmission windows at each altitude are aligned in the table. Hyphens denote windows that do not exist for 90\% transmission but do at 10\% transmission. Curly brackets group multiple windows at 90\% transmission that merge into one window at 10\% transmission. From orbit, we find no 90\% transmission windows.  For 10\% transmission at an altitude of 1 m, only three windows exist. The horizontal lines in the table group which windows at other altitudes ultimately merge to form these three windows. 
\end{landscape}
\newpage

\begin{landscape}
\begin{longtable}{l |  c c c c c c c c c c c c}
\hline
\hline
Species & \multicolumn{12}{|c}{Absorption Bands ($\mu m$)}\\ 
\hline
\hline
& \multicolumn{12}{c}{Amino Acids/Ureas} \\ 
\hline
\hline
\multirow{2}{*}{Glycine$^{1}$}&3.16&3.32&\textbf{3.42}&\textbf{3.49}&3.68&3.83&4.70&5.64&5.87&\textbf{6.28}&\textbf{6.55}&6.83\\&6.93&\textbf{7.07}&\textbf{7.50}&8.83&8.98&9.67\\
\hline
\multirow{3}{*}{Sarcosine$^{1}$}&2.88&\textbf{2.94}&3.14&3.30&3.32&3.37&3.55&3.60&3.95&4.07&\textbf{6.09}&\textbf{6.11}\\&\textbf{6.14}&\textbf{6.30}&6.72&6.83&\textbf{7.09}&\textbf{7.23}&7.63&7.69&7.73&8.55&8.71&9.50\\&9.97\\
\hline
\multirow{2}{*}{$\beta$-alanine$^{1}$}&\textbf{3.32}&\textbf{3.38}&\textbf{3.41}&\textbf{3.44}&\textbf{3.47}&\textbf{3.60}&\textbf{3.79}&\textbf{3.81}&4.53&6.07&\textbf{6.12}&\textbf{6.36}\\&\textbf{6.63}&6.82&6.91&\textbf{7.08}&\textbf{7.12}&\textbf{7.19}&7.45&\textbf{7.50}&\textbf{7.73}&7.98&8.64&9.51\\
\hline
\multirow{2}{*}{$\gamma$-aminobutyric acid$^{1}$}&\textbf{3.42}&\textbf{3.51}&3.73&3.82&4.57&5.25&6.02&6.09&6.25&6.63&\textbf{6.84}&7.00\\&\textbf{7.14}&\textbf{7.23}&\textbf{7.26}&7.47&7.65&7.80&8.03&8.54&8.89&9.43&9.70&9.93\\
\hline
\multirow{2}{*}{Urea$^{1}$}&\textbf{2.91}&\textbf{2.99}&\textbf{3.42}&\textbf{3.49}&\textbf{3.51}&4.97&\textbf{5.96}&\textbf{6.25}&\textbf{6.84}&\textbf{7.26}&7.32&8.69\\&9.98\\
\hline
\hline
\hline
& \multicolumn{12}{c}{Nucleobases} \\ 
\hline
\hline
\multirow{2}{*}{Uracil$^{1}$}&\textbf{3.21}&\textbf{3.22}&\textbf{3.24}&3.29&\textbf{3.35}&3.36&\textbf{3.40}&3.42&3.50&3.54&3.56&3.65\\&5.03&5.66&\textbf{5.74}&\textbf{5.81}&\textbf{5.97}&6.21&6.87&\textbf{7.03}&7.18&\textbf{8.07}&9.10&9.93\\
\hline
\multirow{3}{*}{Purine$^{1}$}&2.91&\textbf{3.22}&\textbf{3.25}&\textbf{3.30}&\textbf{3.32}&\textbf{3.39}&\textbf{3.49}&\textbf{3.59}&\textbf{3.66}&\textbf{3.73}&\textbf{3.74}&\textbf{3.83}\\&\textbf{3.94}&\textbf{6.18}&6.29&\textbf{6.37}&\textbf{6.83}&\textbf{7.00}&\textbf{7.13}&7.37&\textbf{7.50}&\textbf{7.87}&8.13&\textbf{8.23}\\&8.33&\textbf{9.12}\\
\hline
\multirow{2}{*}{Hypoxanthine$^{1}$}&3.19&3.28&3.33&3.35&3.38&3.41&3.46&3.49&3.56&3.62&3.72&3.79\\&3.83&3.76&\textbf{5.95}&6.33&6.81&7.04&7.32&7.41&7.86&8.23&8.68&8.80\\
\hline
\multirow{2}{*}{Xanthine$^{1}$}&2.92&3.19&3.33&3.47&3.53&3.57&5.56&\textbf{5.86}&6.37&6.86&6.95&7.04\\&7.47&7.53&7.56&7.94&8.28&8.35&8.60&8.67&8.92&9.67&9.77\\
\hline
\multirow{2}{*}{Adenine$^{1}$}&3.03&\textbf{3.20}&3.35&3.57&3.69&3.71&3.84&\textbf{5.98}&\textbf{6.23}&6.63&6.80&6.84\\&6.89&7.05&7.31&7.50&\textbf{7.64}&7.92&8.65&8.88&9.78\\
\hline
\multirow{2}{*}{Guanine$^{1}$}&3.01&3.21&3.34&3.44&3.51&3.71&\textbf{5.89}&\textbf{5.97}&6.11&6.39&6.44&6.77\\&6.88&7.05&7.27&7.92&8.22&8.52&8.93\\
\hline
\hline
\hline
& \multicolumn{12}{c}{Alkanes} \\ 
\hline
\hline
\multirow{3}{*}{Methane$^{2}$}&0.89&1.01&1.17&1.41&1.67&1.72&1.79&2.05&2.21&2.21&\textbf{2.32}&2.33$^{w}$\\&\textbf{2.37}&2.38$^{w}$&2.43&\textbf{2.45}&2.60$^{w}$&2.69$^{w}$&2.71$^{w}$&2.76$^{w}$&2.83$^{w}$&\textbf{3.32$^{w}$}&3.44$^{w}$&3.55$^{w}$\\&3.86$^{w}$&\textbf{4.99}$^{*}$&6.26$^{w}$&\textbf{7.70$^{w}$}\\
\hline
\multirow{4}{*}{Ethane$^{2}$}&0.91&1.02&1.19&1.39&1.70&1.73&1.77&\textbf{2.27}&2.27$^{w}$&2.31$^{w}$&\textbf{2.32}&\textbf{2.40}\\&2.40&2.42$^{w}$&2.43&2.46&2.46&\textbf{2.46}&2.65$^{w}$&3.07$^{w}$&3.28$^{w}$&\textbf{3.36$^{w}$}&3.40$^{w}$&\textbf{3.47}\\&3.54$^{w}$&3.66&3.78$^{w}$&3.91&4.24&4.28&4.96&\textbf{4.97}$^{*}$&5.00&6.82&6.86&6.90\\&7.30&7.30\\
\hline
\multirow{3}{*}{Propane$^{3}$}&0.92&1.02&1.19&1.40&1.71&1.76&1.82&2.28&2.32&2.33&2.43&2.47\\&3.14$^{w}$&\textbf{3.39$^{w}$}$^{s}$&\textbf{3.45}&\textbf{3.48}&3.59&3.63&3.65&3.79&4.18&4.87$^{*}$&4.95$^{*}$&6.72\\&\textbf{6.78}&6.82&6.86&7.18&7.24&7.32&7.45&7.54&8.57&8.69&9.34&9.49\\
\hline
\multirow{3}{*}{Butane$^{3}$}&0.91&1.02&1.19&1.40&1.71&1.77&1.82&2.28&2.32&2.34&2.43&2.48\\&3.15$^{w}$&\textbf{3.39$^{w}$}$^{s}$&\textbf{3.47}&3.63&3.77$^{w}$&\textbf{5.06$^{w}$}$^{*}$&6.74&\textbf{6.78}&\textbf{6.81}&\textbf{6.83}&7.15&7.20\\&7.41&7.45&7.66&7.70&7.76\\
\hline
\multirow{3}{*}{Pentane$^{3}$}&0.91&0.94&1.02&1.19&1.22&1.39&1.73&1.77&1.83&2.28&\textbf{2.31}&2.35\\&2.44&2.47&3.13$^{w}$&\textbf{3.42$^{w}$}$^{s}$&3.65$^{w}$&3.75&3.81&4.86$^{*}$&4.94$^{*}$&5.01$^{*}$&\textbf{6.86$^{w}$}$^{s}$&\textbf{7.25}\\&7.47&7.67&7.94&8.10&8.83$^{w}$&9.40&9.78\\
\hline
\multirow{3}{*}{Hexane$^{3}$}&0.91&0.94&1.03&1.19&1.22&1.40&1.73&1.77&1.83&2.28&\textbf{2.31}&\textbf{2.35}\\&2.36&2.44&2.47&\textbf{3.44$^{w}$}$^{s}$&3.66&3.74&3.83&4.94$^{*}$&5.00$^{*}$&\textbf{6.84$^{w}$}$^{s}$&\textbf{7.24$^{w}$}&7.44\\&7.64&7.72&8.00&8.19&8.80&9.38&9.66&9.87\\
\hline
\multirow{1}{*}{Cyclohexane$^{3}$}&3.17$^{w}$&\textbf{3.41$^{w}$}&\textbf{3.50}&3.58&3.72&3.75&3.76&3.83&6.87&7.95&9.62&9.90\\
\hline
\hline
\hline
& \multicolumn{12}{c}{Alkenes/Alkynes} \\ 
\hline
\hline
\multirow{3}{*}{Acetylene$^{2}$}&1.06&1.19&1.36&1.40&1.42&1.49&1.51&1.53&1.54&1.55&1.57&1.58\\&1.86&1.90&1.93&1.93&1.95&2.00&2.04&2.21&2.45$^{w}$&2.59&\textbf{3.10$^{w}$}&4.85$^{w*}$\\&4.94$^{*}$&5.85$^{w}$&7.19\\
\hline
\multirow{2}{*}{Ethylene$^{2}$}&2.13&2.19&2.22&2.26$^{w}$&2.32&2.32$^{w}$&2.34&2.39&3.24&3.26&3.36&3.37\\&4.12&4.61$^{w}$&4.90$^{*}$&4.90&5.08$^{*}$&5.08&5.26$^{w}$&6.94&6.96&9.60&9.65\\
\hline
\multirow{1}{*}{Propene$^{3}$}&\textbf{3.22}&\textbf{3.25}&\textbf{3.38$^{w}$}&3.50&5.44&5.51&\textbf{6.00}&6.05&6.11&6.78&\textbf{6.92}&9.56\\
\hline
\hline
\hline
& \multicolumn{12}{c}{Nitriles} \\ 
\hline
\hline
\multirow{1}{*}{Hydrogen Cyanide$^{2}$}&2.38$^{w}$&\textbf{3.19$^{w}$}&4.27&4.42$^{w}$&\textbf{4.76}&4.84$^{*}$&4.84&5.02$^{*}$&6.17$^{w}$\\
\hline
\multirow{3}{*}{Acetonitrile$^{2}$}&2.22&2.25&2.27&2.29&2.48$^{w}$&2.52&3.12&3.16&3.33&3.40&3.65&3.80\\&4.09&4.15&4.36&4.44&6.88&7.04&7.06&7.10&7.26&7.29&9.54&9.62\\&9.65\\
\hline
\multirow{1}{*}{Cyanogen$^{2}$}&3.74&4.27&4.62\\
\hline
\multirow{2}{*}{Acrylonitrile$^{1}$}&2.71&3.23&3.26&3.30&4.39&4.48&5.22&6.07&6.21&6.23&\textbf{7.09}&7.33\\&7.78&9.16\\
\hline
\hline
\hline
& \multicolumn{12}{c}{Aromatic Organic Molecules} \\ 
\hline
\hline
\multirow{1}{*}{Benzene$^{3}$}&\textbf{3.23}&\textbf{3.25}&\textbf{3.29}&4.97$^{*}$&\textbf{5.05}$^{*}$&5.10&5.51&6.57&\textbf{6.76}&7.20&8.55&\textbf{9.68}\\
\hline
\multirow{3}{*}{Naphthalene$^{3}$}&3.23&\textbf{3.25}&\textbf{3.27}&3.35&3.37&4.92$^{*}$&5.03$^{*}$&5.14&5.23&5.27&5.41&5.46\\&5.53&5.64&5.81&5.99&6.04&6.27&6.46&6.57&\textbf{6.62}&7.20&7.36&7.89\\&8.09&8.26&8.78&8.87&\textbf{9.87}\\
\hline
\multirow{3}{*}{Biphenyl$^{1}$}&3.21&3.24&3.26&3.28&3.29&3.37&3.38&3.41&3.42&3.44&3.48&5.14\\&5.33&6.25&6.37&\textbf{6.75}&6.99&7.22&7.26&7.43&7.64&8.45&8.55&8.66\\&8.99&9.16&9.29&9.59&9.93\\
\hline
\multirow{3}{*}{Indole$^{1}$}&2.57&2.86&\textbf{2.94}&3.28&\textbf{3.39}&\textbf{3.42}&\textbf{3.49}&6.19&6.34&6.64&6.72&\textbf{6.87}\\&7.05&7.26&7.32&7.39&7.47&7.82&8.01&8.29&8.93&9.25&9.43&9.79\\&9.89&9.95\\
\hline
\hline
\hline
& \multicolumn{12}{c}{Inorganics} \\ 
\hline
\hline
\multirow{1}{*}{Water$^{4}$}&\textbf{3.04$^{w}$}&4.55$^{w}$&6.04$^{w}$\\
\hline
\multirow{1}{*}{Ammonia$^{4}$}&2.00$^{w}$&2.23$^{w}$&2.96$^{w}$&3.11$^{w}$&6.15$^{w}$&9.35$^{w}$\\
\hline
\multirow{1}{*}{Carbon Monoxide$^{4}$}&2.34&2.35&2.65&2.66&2.69&2.70&\textbf{4.68}&4.78&6.16&6.22&6.24&6.28\\
\hline
\multirow{1}{*}{Carbon Dioxide$^{4}$}&2.70&2.78&\textbf{4.26$^{w}$}&4.38\\
\hline
\end{longtable}

\textbf{Table 2}. Absorption bands for species of geomorphologic, chemical, and prebiotic interest. Absorption databases are referenced for each compound. Features with $<$30\% transmission from their respective database are considered ``strong" and are included in bold.  Features either identified in the literature as wide or having a full width at 90\% peak of $>$10 cm$^{-1}$ are considered ``wide" and labelled with ``w".  Saturated features are labelled with ``s".  Finally, additional $\sim$5.0 $\mu$m absorptions taken from Clark et al. 2010 are labelled with ``*". For C$_{2}$H$_{2}$, additional absorptions with wavenumber $>$4000 cm$^{-1}$ ($<$2.5$\mu$m) are provided from Singh et al. 2015.  For all alkanes, additional absorptions with wavenumber $>$4000 cm$^{-1}$ ($<$2.5$\mu$m) are provided from Clark et al. 2009. References: 1. Spectral Database for Organic Compounds  2. Cosmic Ice Laboratory  3. NIST Chemistry WebBook 4. Ice Analogs Database.  See the text for more details on the conditions in which each spectrum is calculated.
\end{landscape}

\newpage

\section*{Figures}

\centerline{\includegraphics[width=0.9\linewidth]{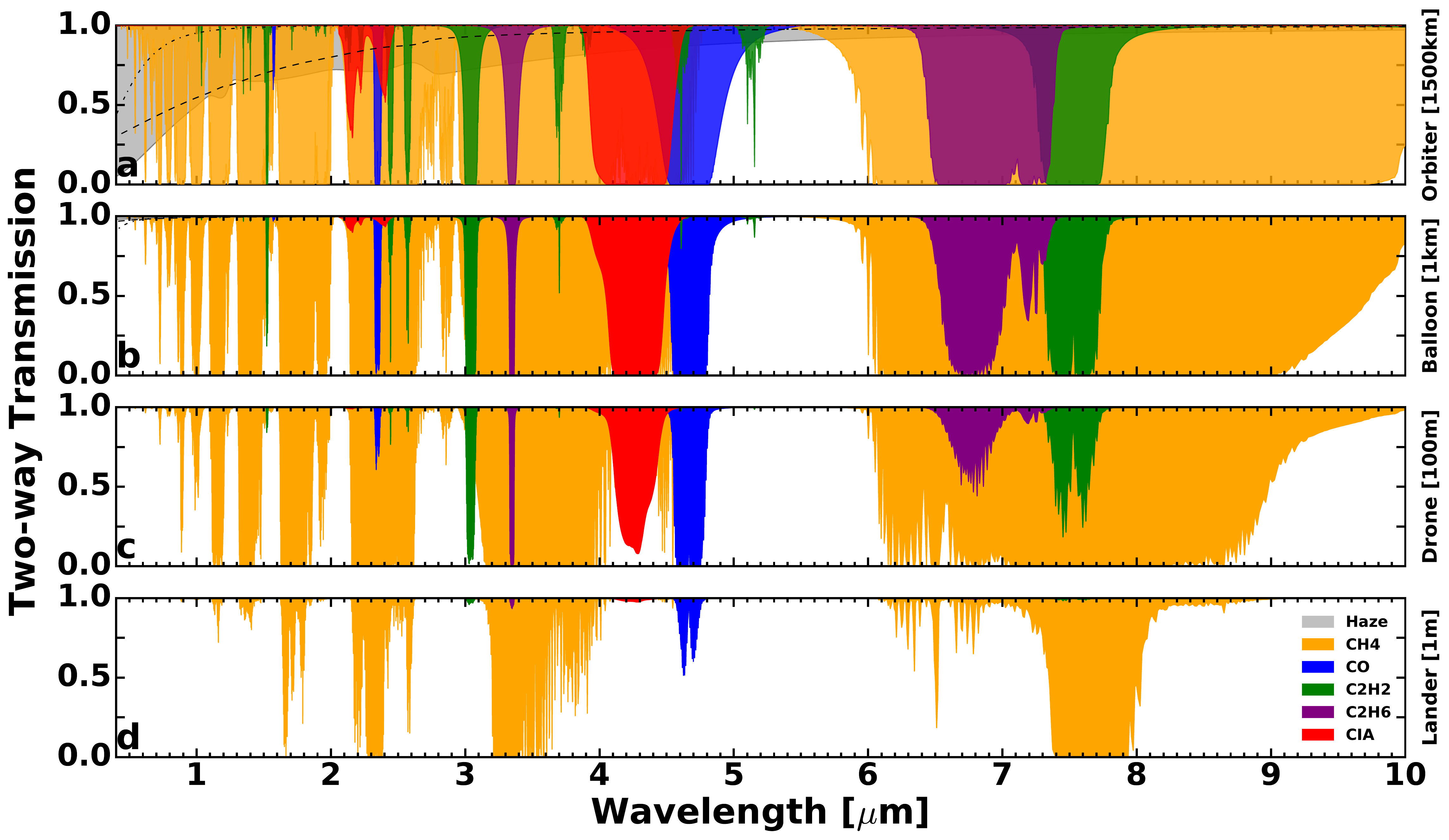}}
\textbf{Fig. 1}. The two-way gaseous absorption in Titan's atmosphere for four different nominal mission architectures: orbiter, balloon, drone, and lander, as shown to the right of each panel.  White areas indicate spectral regions with signal above zero percent.  Methane (orange) is the primary absorber in Titan's atmosphere.  Other absorption from trace gases further contribute to the transmission.  Contributions from Rayleigh and haze scattering (were they to be treated independently of each other and absorption) are included as dotted and dashed lines, respectively.  From orbit, only very narrow band windows are found that can see through to Titan's surface.  The 5.0 $\mu$m window is the widest and most transparent.  We note that the absorption at 1500 km are comparable to 1000 km, the lowest altitude achieved by Cassini, and is also representative of future "close-flyby" mission architectures.  As the mission altitude decreases, absorption in the windows decreases and the wings of the windows widen to allow for greater spectral coverage.  By the surface, most of Titan's near-IR spectrum is observable, barring a few narrow regions of exceptionally strong methane absorption.
\label{Fig1}
\newpage

\centerline{\includegraphics[width=0.9\linewidth]{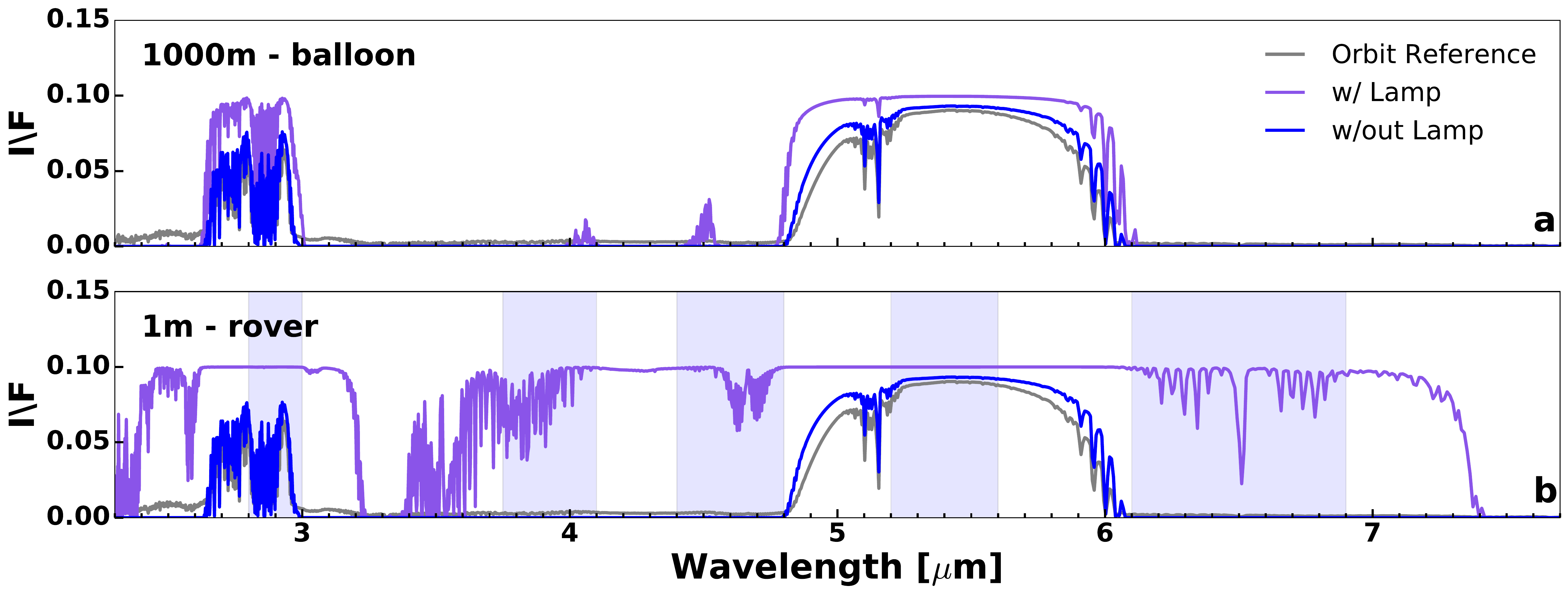}}
\textbf{Fig. 2}. Comparison of Titan's simulated apparent reflectance (I/F) between passive and active illumination at two different nominal altitudes.  Spectra are plotted from 2.5 - 7.5 $\mu$m as the range of greatest interest for absorption features (see Table2/Figure 3).  The orbital reference I/F (grey) matches closely to the ''without lamp" cases (blue) because of the significant one-way opacity of Titan's atmosphere that effectively acts to extinct all incident solar light, except for in the windows.  The orbital I/F, a model spectrum at the top of the atmosphere, is typically higher than the nominal mission case because of contributions from scattering in the upper atmosphere.  \textbf{a}: comparison at 1km.  Even if strong enough broadband onboard illumination existed for such an altitude only a modest improvement is observed in spectral coverage on the wings of each window (purple), suggesting a maximum mission altitude with which to obtain greater spectral coverage over an orbiter. \textbf{b}: comparison at 1 m.  Significant enhancement in the spectrum is observed, with an increase in window width as high as $\sim$200\%.  Low altitude missions, then are favored for major improvements in spectral coverage in the near-IR, but they must have an active light source, otherwise there is no significant improvement.  In this panel, shaded areas demonstrate typical coverage available from commercial LEDs are also indicated \citep{Boston2018}.  Spectral enhancement would be limited to these regions only for LED sources.  See the text in section \ref{res:passive_active} for a more detailed discussion of using LEDs as a source of active illumination. Because of this spectral broadening, several LED's are required to cover the entire windows in the mid-IR, i.e. we define 4 LED bands in the 5.0 $\mu$m window and have only covered $\sim$50\% of the entire window.  
\label{Fig2}
\newpage

\centerline{\includegraphics[width=0.9\linewidth]{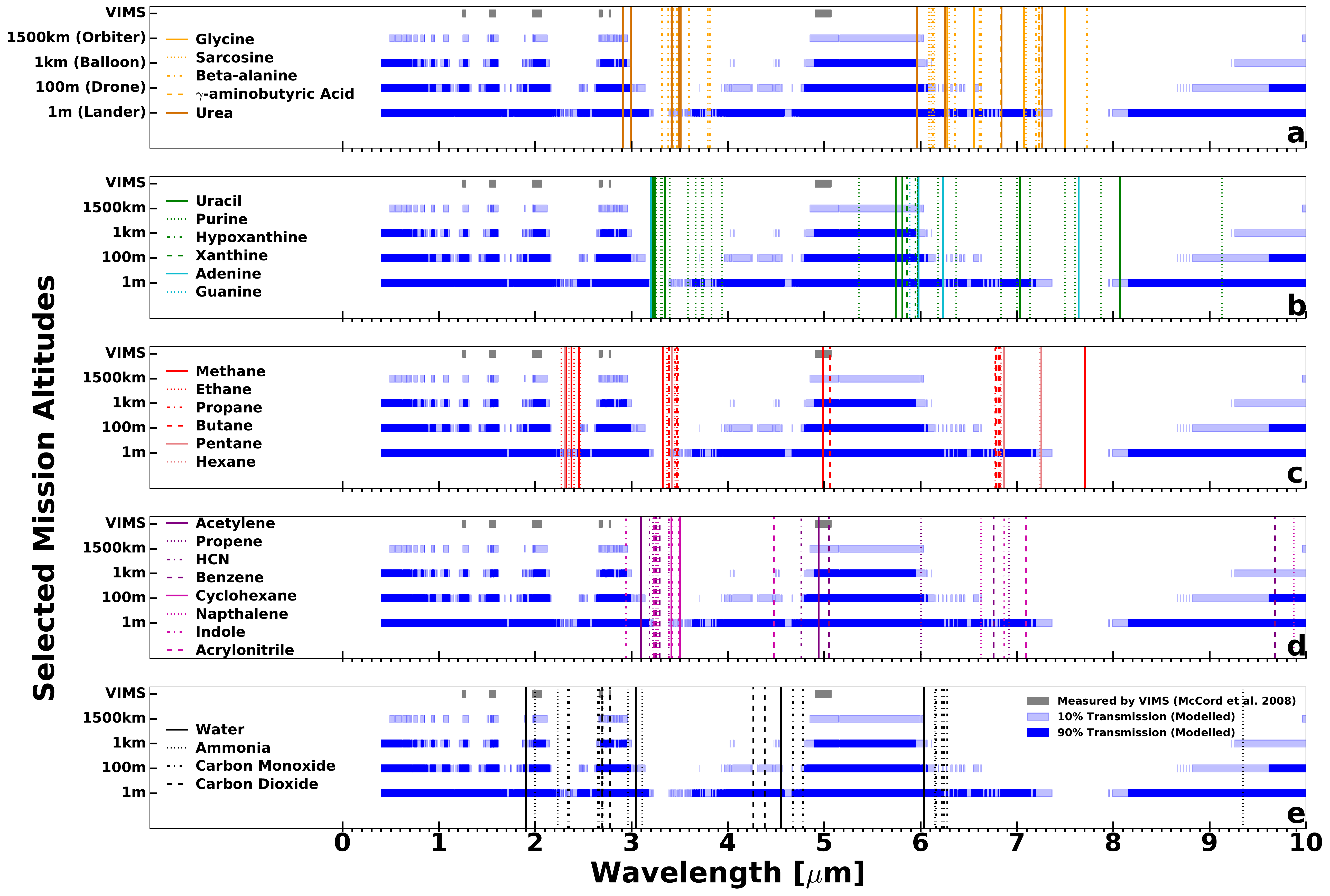}}

\textbf{Fig. 3}. The spectral windows available from orbit as defined by \citealt{McCord2008} for VIMS, as modeled from orbit (1500 km), and at altitudes relevant for future balloon (1 km), drone (100 m), and lander (1 m) mission architectures assuming the use of active onboard illumination. The subpanels are divided based on the types of potential surface compounds shown: \textbf{at)} Amino acids/amides \textbf{b)} nucleobases \textbf{c)} alkanes \textbf{d)} nitriles, alkenes and aromatic hydrocarbons \textbf{e)} inorganics. The windows are defined using the horizontal shaded boxes as regions where the atmospheric transmission is between 90\% (minimum width) and 10\% (maximum width). The vertical lines denote major absorption bands for surface compounds of geological, chemical or prebiotic interest, and weaker lines that are of particular use in the windows.  Lines at wavelengths $<$2.0 $\mu$m are not plotted because of either lack of data for this region or because of difficulties in discerning between similar and multiple weak overtone features of compounds within a given family. Wavelengths especially $>$ 3.0 $\mu$m are more favored due to the strength and uniqueness of the C-H absorption features in this region. For more details on line strengths, widths, and lines not plotted here, please see Table 2.

\label{Fig3}
\newpage

\section*{References}
\bibliographystyle{elsarticle-harv}
\bibliography{library}

\newpage
\newgeometry{left=4.5cm, right=4.5cm, bottom=4.5cm, top=4.5cm}

\section{Supplementary Material}
\subsection{Model Validation using VIMS Observations}

Here we demonstrate that the radiative transfer model used to report the atmospheric windows in the lower atmosphere can fit and replicate VIMS observations of the Huygens landing site.  The intent of this Supplementary Material is to offer more background on how certain atmospheric inputs to the model are derived and to provide a comparison of the model to VIMS observations to provide the reader with a level of confidence in the output of our model.  When possible, inputs to the model are derived from Huygens landing site observations, so as to provide a reasonable estimate of the properties of Titan's atmosphere (see Section 2 of the main text for details).  We note, however, that atmosphere circulation is know to vary (e.g. seasonal circulation) and so too will atmospheric properties (e.g. vertical haze abundance, methane surface humidity, etc.), but here we choose the Huygens landing profile, as our only ground truth of Titan's atmosphere, for this study.

As alluded to in Section 2.1.2 of the main text, single scattering albedo (SSA) remains an important input for modelling the scattering of Titan's atmosphere and opacity of the hazes.  In order to constrain the single scattering albedo in the near-IR we use a similar methodology of \citep{Hirtzig2013}.  Therein, a study was performed to minimize the 2-D chi-squared space of single scattering albedo vs vertical haze abundance as a function of wavelength for the Huygens landing site, which was approximated by a linear combination of pixels [11,3],[11,4],[12,3],[12,4] with weights of 0.625, 0.125, 0.125, 0.125 respectively, which were averaged together from cube CM1481624349 \citep{Hirtzig2013}.  They found that the nominal haze abundance profile, as measured by the Huygens probe, was applicable for all wavelengths, provided a wavelength-dependent single scattering albedo was adopted.  Here, we perform a similar study, using the same combination of pixels to approximate the Huygens landing site, but updating the haze profiles and scattering phase functions to those described in \cite{Doose2016}.

First, a model is setup with the nominal profiles for haze \citep{Doose2016} and methane \citep{Niemann2010} abundances as derived from Huygens.  Then, the single scattering albedo is varied uniformly to generate a family of spectra with which to compare against the Huygens landing site observation.  These spectra are inverted for each wavelength in order to derive the required wavelength-dependent single scattering albedo that best matches the observation. Lastly, we interpolate between a running average of the channels in the ``wings" of each spectral window, in order to determine the final single scattering albedo as a function of wavelength.  We choose to only interpolate the ``wing" channels to remove the possible influence of surface albedo (surface channels) or high methane opacity (band channels) on the fit.  The derived single scattering albedo is plotted in \textbf{Fig. S1}.  We find overall good agreement between our newly derived single scattering albedo using the updated haze scattering properties and that derived in Hirtzig et al. (2013), with two primary difference.  First, is a decrease in the single scattering albedo at shorter wavelengths, which is now needed in order to match the observations. Second, we adopt a less variable SSA albedo for wavelengths longer than 3$\mu$m.  Because atmospheric opacity is so high in this region,  the SSA is poorly constrained, leading to little support of any derived variations in SSA in this spectral region. Regardless, variations in the SSA in regions of such strong absorption have little influence on observed spectra because of how quickly light is absorbed out of the beam, minimizing contributions from scattering.  The single scattering albedo we derived here, as shown in \textbf{Fig. S1}, we use for wavelengths $>$1.0 $\mu$m.  For wavelengths $<$1.0 $\mu$m we adopt the altitude-dependent single scattering albedo as derived in \cite{Doose2016}, which are derived from the Huygens DISR data.

In order to verify that the use of the \cite{Doose2016} SSA in concert with our model is able to match existing observations of Titan through the near-IR, in \textbf{Fig. S2} we compare our model output with a VIMS reflectance spectrum of the Huygens landing site. We find that by applying $<$ 5\% variations to the \cite{Doose2016} SSA and haze vertical abundance, we are able to match the observations to within error (within the atmospheric bands that are sensitive to Titan's hazes) down to the 0.88 $\mu m$ limit of the VIMS spectrum.  To fit the windows, derivation of the surface albedo is also required.  For this, we employ a similar technique used for deriving the SSA, where a series of models are run with varying surface albedos that are then inverted to find the best match surface albedo spectrum to the Huygens landing site data.  Because we are able to reproduce observations down to 0.88 $\mu m$, and the \cite{Doose2016} SSA and KT10 band model are the best available parameters for visible wavelengths down to our 0.4 $\mu m$ limit, we use the as the inputs for our derivation of the windows at visible wavelengths.  However, we note that while a robust comparison to Huygens data is not provided here, and emphasize again our focus on the infrared windows in terms of highest accuracy (see Section 3.1 of main text). We find overall good agreement between the model and the observations, except for the spectral region from $\sim$2.9-3.1 $\mu m$, which is a spectral region in which all line lists at Titan like-conditions are incomplete.  Future laboratory measurements will be needed to more accurately model this spectral region.  However, this comparison demonstrates that in addition to atmospheric gaseous absorption we are able to fit physical parameters related to scattering of Titan's hazes and surface, thus validating the major components of this model.  This demonstrates the validity of the model over the broad spectral region of the near-IR and further complements the validation of the model previous conducted in other studies \cite{Adamkovics2016,Adamkovics2017,Lora2017}.  

\subsection{Comparison of correlated k-coefficients}

PyDISORT uses correlated k-coefficients to allow for the rapid modelling of the thousands of individual methane absorption lines that populate Titan visible to near-IR spectrum.  Correlated k-coefficients are generated from high-resolution line-by-line calculations of the absorption strength of methane over a uniform grid of temperatures and pressures, which are then interpolated to the appropriate conditions of Titan's atmosphere.  The line lists used for these calculations can significantly impact the model. We discuss here three line lists, two of which we use to accurately model the transmission in Titan's atmosphere. 

The first of these line lists is from HITRAN 2016 \cite{Gordon2017}.  Over the years, this database has been regularly updated with improved laboratory and modelling efforts and has been extensively used in radiative transfer models of Titan's atmosphere  \cite{Campargue2012,Griffith2012,Hirtzig2013,Adamkovics2016}.  Leveraging the combination of empirical models and laboratory data, HITRAN 2016 offers excellent characterization of methane lines for wavelengths $>$1.4$\mu$m.

The second line list is the Theoretical Reims-Tomsk Spectral (TheoReTS) database from the recent work of \cite{Rey2017}.  This database makes use of ab initio calculations to more completely determine the absorptions in the overtones of methane for wavelengths down to 0.75$\mu$m, which the empirical models of HITRAN 2016 fail to accurately reproduce.  \cite{Rey2017} report discrepencies between the integrated intensities of the two lines lists to be as high as 10\% for wavelengths ranging from 0.96$\mu$m-1.05$\mu$m (see Table 1 of \cite{Rey2017}). 

The third list is the empirically derived model of methane opacity from \cite{Karkoschka2010}.  Therein, they use a combination of laboratory spectra, Huygens DISR observations, and Hubble Space Telescope observations of Jupiter to derive k-coefficients for methane that extend down into the visible.  These lines have been extensively used in the literature and have been demonstrated to accurately model Titan's atmosphere over a wide range of wavelengths in the near-IR \cite{DeBergh2012,Campargue2012,Hirtzig2013,Rey2017}.

\cite{Rey2017} found good agreement between these three line lists for Titan from orbit, with the HITRAN line list in good agreement for wavelengths $>$1.4$\mu$m and the KT10 line list in good agreement at all wavelengths down to 0.75$\mu$m. We performed an additional comparison of these three line lists by simulating spectra of Titan's atmosphere in the lowest 1-km (see \textbf{Fig. S3}). We found good agreement using the KT10 line list $<$1.4 $\mu$m and HITRAN $>$1.4$\mu$m, using the cutoff value as suggested from the results of \cite{Rey2017}.  Use of the KT10 line list offers the further benefit of being able to consider visible wavelengths in this work.  For these reasons, we opted for a combination of the KT10 and HITRAN 2016 line lists here.

\subsection{Figures}

\centerline{\includegraphics[width=1\linewidth]{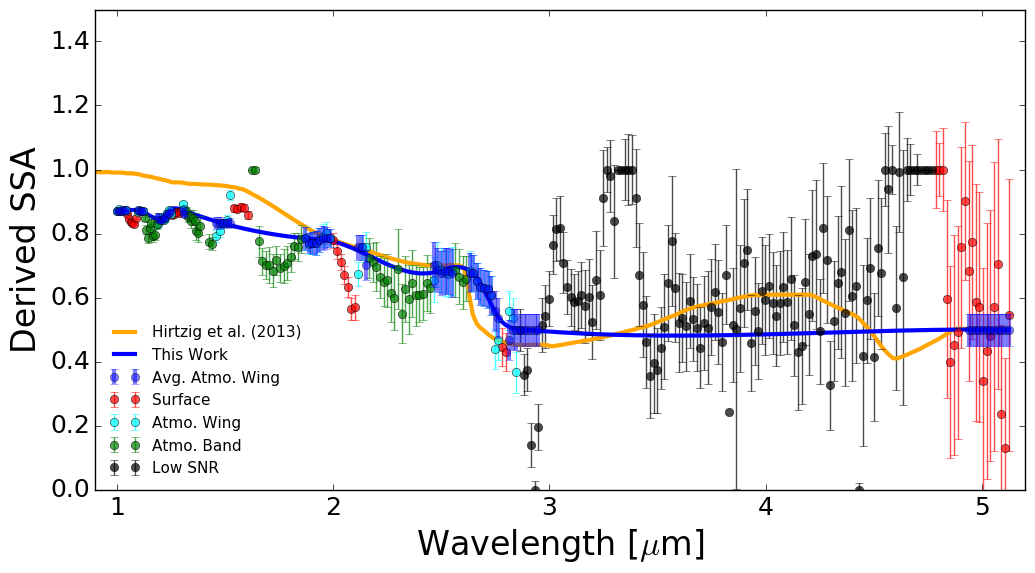}}
\label{fig s1}
\textbf{Fig. S1}. The derived single scattering albedo for Titan's hazes using a VIMS observation of the Huygens Landing Site.  Channels corresponding to Titan's surface are plotted in red, to the lower atmosphere in cyan, and to the upper atmosphere in green.  Point in black are in regions of high methane opacity, and therefore insufficient SNR for accurate retrievals.  In blue is a 5-point running average of the cyan points, which is interpolated to determine the single scattering albedo vs wavelength. In orange is the single scattering albedo as derived in Hirtzig et al. (2013).  We find overall good agreement between the two vectors, except at shorter wavelengths, where a darker single scattering albedo is needed.

\centerline{\includegraphics[width=1\linewidth]{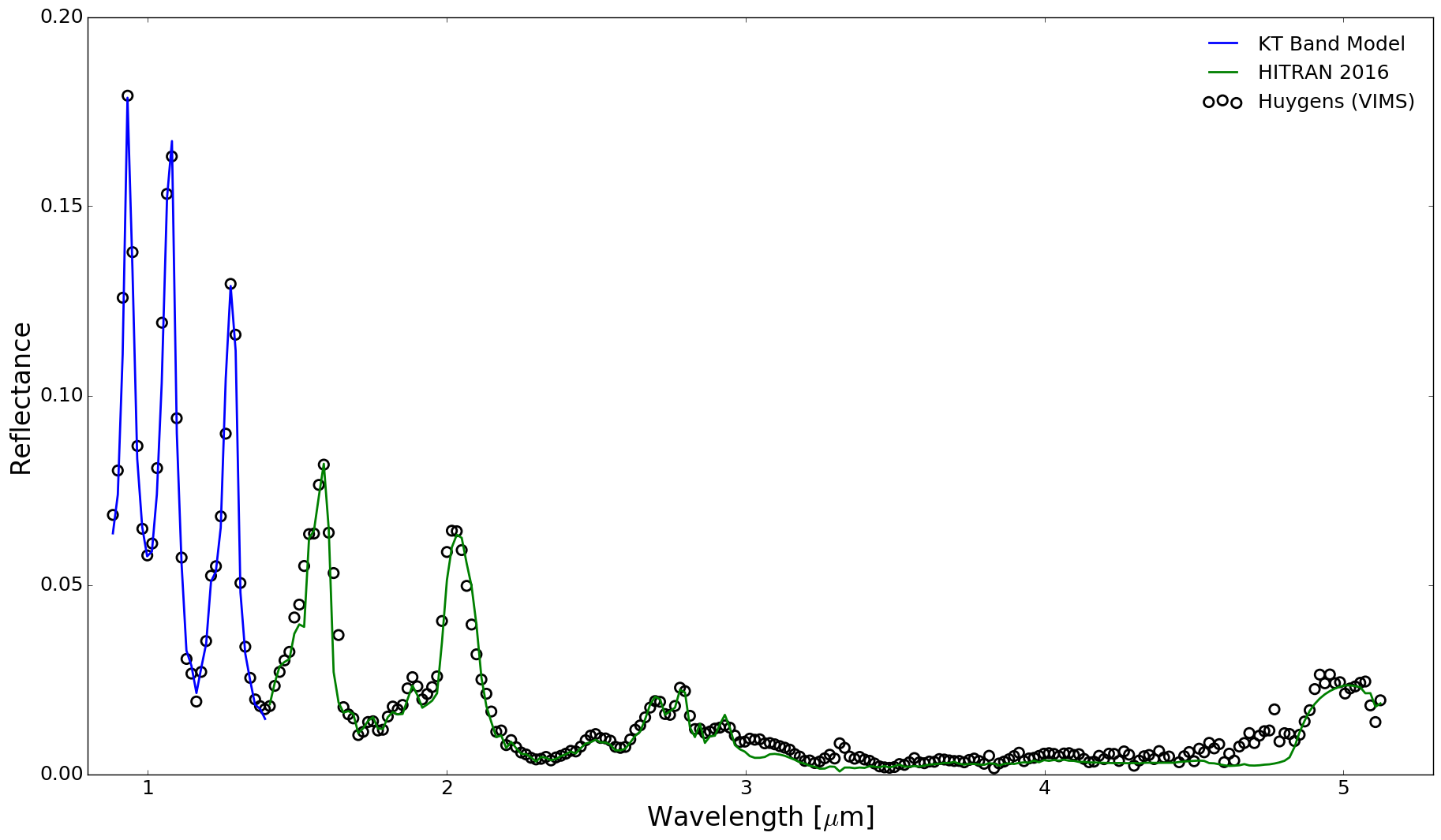}}
\label{fig s2}
\textbf{Fig. S2}. A fit of the radiative transfer model to the reflectance of a VIMS observation of the Huygens landing site.  Using the derived single scattering albedo and surface albedo, we find good overall agreement between the model and the VIMS observations.  The region from 2.9-3.2 $\mu$m is imperfectly fit because of the incompleteness in all line lists at these wavelengths.  Future efforts will be needed to model the absorption of both methane, as well as other possible atmospheric absorbers at these wavelengths.

\newpage

\centerline{\includegraphics[width=1\linewidth]{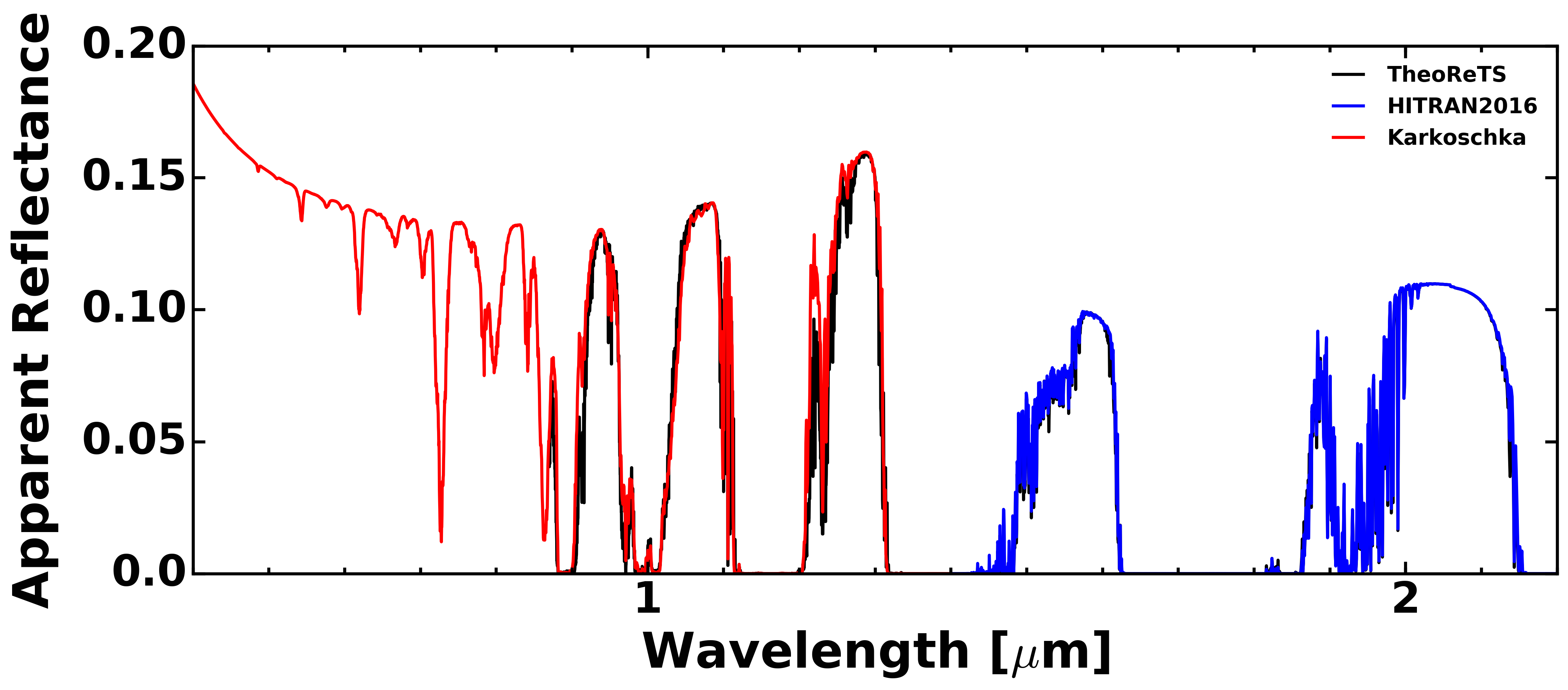}}
\label{fig s3}
\textbf{Fig. S3}. A comparison of three separate methane line lists at 1km altitude.  Line lists from TheoReTS \cite{Rey2017} are plotted in black, HITRAN 2016 \cite{Gordon2017} in blue, and Karkoschka and Tomasko 2010. in red.  We find good agreement between the line lists over their respective regions of interest, as also described in \cite{Rey2017}.

\label{huygens_fit}

\end{document}